# LED-based Photo-CIDNP Hyperpolarization Enables $^{19}$F MR Imaging and $^{19}$F NMR Spectroscopy of 3-fluoro-DL-tyrosine at 0.6 T

Johannes Bernarding[1*], Christian Bruns[1], Isabell Prediger[1], Markus Plaumann[1]

## Abstract

Although $^{19}$F has high potential to serve as a background-free molecular marker in bioimaging, the molar amount of marker substance is often too small to enable $^{19}$F MR imaging or $^{19}$F NMR spectroscopy with a sufficiently high signal-to-noise ratio (SNR). Hyperpolarization methods such as parahydrogen-based hyperpolarization or dynamic nuclear polarization (DNP) can significantly improve the SNR, but require expensive and complex sample preparation and the removal of toxic catalysts and solvents. Therefore, we used the biologically compatible model of the fluorinated amino acid 3-fluoro-DL-tyrosine with riboflavin 5′-monophosphate (FMN) as a chromophore dissolved in $D_2O$ with 3.4% $H_2O_{dest.}$ allowing to transform light energy into hyperpolarization of the $^{19}$F nucleus via photo-chemically induced dynamic nuclear polarization (photo-CIDNP). We used a low-cost high-power blue LED to illuminate the sample replacing traditionally used laser excitation, which is both potentially harmful and costly.

For the first time, we present results of hyperpolarized $^{19}$F MRI and $^{19}$F NMR performed with a low-cost 0.6 T benchtop MRI system. The device allowed simultaneous dual channel $^1$H/$^{19}$F NMR. $^{19}$F imaging was performed with a (0.94 mm)$^2$ in-plane resolution. This enabled the spatial resolution of different degrees of hyperpolarization within the sample. We estimated the photo-CIDNP-based $^{19}$F signal enhancement at 0.6 T to be approximately 465. FMN did not bleach out even after multiple excitations, so that the signal-to-noise ratio could be further improved by averaging hyperpolarized signals. The results show that the easy-to-use experimental setup has a high potential to serve as an efficient preclinical tool for hyperpolarization studies in bioimaging.

## Abbreviations

| | |
|---|---|
| CIDNP | chemically induced dynamic nuclear polarization |
| DNP | dynamic nuclear polarization |
| FID | free induction signal |
| FMN | riboflavin 5′-monophosphate |
| ISC | intersystem crossing |
| LF | Larmor frequency |
| MRI | magnetic resonance imaging |
| NMR | nuclear magnetic resonance |
| PHIP | parahydrogen induced polarization |
| SABRE | signal amplification by reversible exchange |
| SCRP | spin-correlated radical pair |
| SE | signal enhancement |
| SI | signal intensity |
| SNR | signal-to-noise ratio |
| TE | echo time |
| TFE | trifluoroethanol |
| TR | repetition time |
| ZF | zero filling factor |

## Affiliations

[1*]Institute for Biometry and Medical Informatics, Medical Imaging Lab, Medical Faculty of the Otto-von-Guericke University Magdeburg, D-39120 Magdeburg, Leipziger Strasse 44, Germany

*Corresponding author: Johannes Bernarding, johannes.bernarding@med.ovgu.de,
ORCID: 0000-0001-6429-4851

## Acknowledgement

We thank Dr. Jürgen Braun (Charité-Universitätsmedizin Berlin, Institute of Medical Informatics, Germany) for helpful discussions. We also thank the team of Pure Devices GmbH (Würzburg, Germany) for their steady and fast support as well as for the helpful joint discussions.

# 1 Introduction

The high potential of $^{19}$F to serve as a background-free molecular marker in biological samples is often hampered by the low concentrations of $^{19}$F-based biomarkers, resulting in a signal-to-noise ratio (SNR) that is often too low for direct observation of $^{19}$F [1–3]. Hyperpolarization of the $^{19}$F nucleus in organic molecules offers a promising strategy for increasing the SNR. All hyperpolarization techniques can serve to transfer non-Boltzmann polarization to heteronuclei [4–6] but known techniques such as *dynamic nuclear polarization* (*DNP*) or parahydrogen-based hyperpolarization techniques such as *parahydrogen induced polarization* (*PHIP*) or *signal amplification by reversible exchange* (*SABRE*) not only require expensive experimental setups for the generation of hyperpolarized substrates, but also considerable efforts to remove the toxic solvents and catalysts inherent in these techniques. This complicates the routine application of these techniques for bioimaging.

To overcome this limitation, a third, long-known alternative can be used since it enables the generation of hyperpolarization in biocompatible model systems. The so-called *photo-chemically induced dynamic nuclear polarization* (*photo-CIDNP*) effect [7, 8] serves usually to analyze reactions involving radicals [9]. Here we used photo-CIDNP to generate hyperpolarization of the $^{19}$F nucleus by illuminating a flavin derivative that acts as chromophore, which leads to a non-equilibrium enrichment of nuclear spins in the involved chemical products via a radical pair mechanism (Fig. 1; for details see Kuprov et al. [10, 11]).

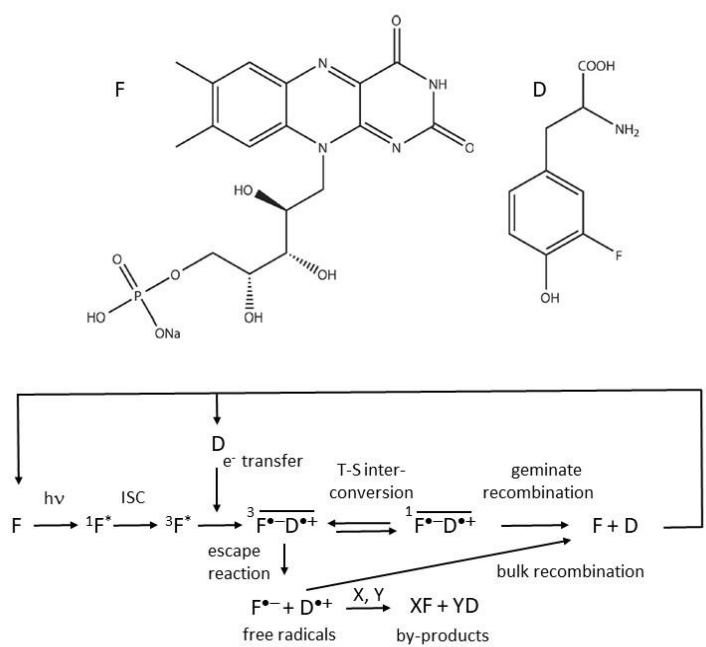

**Fig. 1** Simplified reaction pathways in liquid-state photo-CIDNP of the riboflavin 5'-monophosphat/3-fluoro-DL-tyrosine system under medium- and high-field conditions.
Light irradiation (hν) induces an excited singlet state in the chromophore F (here: riboflavin 5′-monophosphate) that undergoes intersystem crossing (ISC) into a molecular triplet state. This triplet state is highly reactive attracting an electron from the donor D (here: 3-fluoro-DL-tyrosine) and forming a spin-correlated radical pair (SCRP) under conservation of the total spin multiplicity. The SCRP in its triplet state reacts further on two competing pathways. On the first pathway, the triplet state undergoes a transition to a singlet state via coherent triplet-singlet interconversion due to the different Larmor frequencies of the two electrons and oscillates between triplet and singlet state. Hyperfine interaction and the difference of the Landé factors g of the electrons within the two radicals (Δg, [11]) lead to spin-sorting of the nuclear spin states, which finally gives rise to a non-equilibrium enrichment of the nuclear spins in the different chemical products. The subsequent recombination (geminate recombination) into the original educts can only take place when the SCRP is in its singlet state. In the second pathway, the molecules diffuse apart as free radicals (escape products) from the solvent cage into the surrounding solvent where they recombine by either forming by-products with other molecules X, Y or by recombining into the original educts (bulk recombination). According to Kuprov et al. [10, 11] only the geminate contribution accounts for the observable steady state $^{19}$F photo-CIDNP effect in 3-fluoro-D/L-tyrosine as the spin-lattice relaxation time of $^{19}$F nuclei in the free radicals is very fast.

If no by-products are formed, both reaction pathways will result in the primary reactants, and the sample can be irradiated multiple times until inevitable chromophore bleaching and potential photo-generated products reduce the efficiency of the reaction. In this way, data averaging can be used to further increase SNR, while other hyperpolarization techniques are either limited to a single data acquisition (e.g. DNP) or require the addition of new hydrogen during measurement cycles (e.g. PHIP and SABRE).

Here, we investigated the $^{19}$F hyperpolarization of the amino-acid 3-fluoro-DL-tyrosine in an aqueous solution of $D_2O$ with added $H_2O_{dest.}$ (see Materials and Methods section) using riboflavin 5′-monophosphate as the chromophore. Since non-fluorinated tyrosine is an important amino acid in many living organisms, and flavins play an important role in various biological radical reaction schemes [12], this model system can be considered biocompatible since $^{1}$H is also replaced by $^{19}$F in many important fluorinated drugs for medical use in humans [13–15].

Hyperpolarized 3-fluoro-DL-tyrosine has been studied in detail both experimentally and theoretically by Kuprov et. al. [16, 17]. They used an argon laser and optimized the glass fiber transmitting the light into the sample to increase the uniform irradiation of the sample and thus the achievable signal.

$^{1}$H photo-CIDNP of amino acids and other molecules is a well-established technique for many years [18] and was used early on to study amino-acids, as they can be used as probes to monitor structure changes and kinetics in proteins [19–22]. To analyze the field dependence of photo-CIDNP for various amino acids and other molecules, the group of Vieth et al. has developed a special technique to generate 1H photo-CIDNP over a wide range from near zero T to 7T [23, 24]. Field dependency was also investigated by [25]. Lyon et al. used $^{1}$H photo-CIDNP to analyze the structures of disordered, partially folded states of proteins using individual tryptophan and tyrosine residues [26], while Khan et al. applied $^{19}$F NMR to investigate the folded and native structure of green fluorescent protein [22]. Kiryutin et al. showed that $^{1}$H photo-CIDNP can serve for creating long-lived spin order in biologically relevant molecules [27]. Sheberstov et al. investigated long-living heteronuclear singlet order ($^{1}$H, $^{13}$C) generated by of photo-CIDNP at earth magnetic field strength [28]. $^{1}$H-detected $^{13}$C photo-CIDNP can serve as a sensitivity enhancement tool for liquid solution NMR [6, 29]. Solid state $^{1}$H, $^{13}$C, and $^{15}$N photo-CIDNP effect can be used to analyze the amino acid tryptophan [30], and served in a new approach using magic angle spinning NMR to analyze photochemical processes in bacterial reaction centers [31–33]. Matysik et al. [33] provided a detailed overview of how biological solid-state NMR photo-CIDNP enables the study of the dynamics of nuclear spins in photosynthetic reaction centers of bacteria, algae, and higher plants.

All these studies were spectroscopic studies where experimental conditions are quite different to data acquisition in *Magnetic Resonance Imaging (MRI)* which is usually performed with lower bandwidth at the cost of strong reduction or complete loss of the spectral information. In standard clinical MR scanners the resolution is typically one to several mm$^2$ in-plane. Signal and contrast are functions of the amount of NMR-active isotopes of the studied nuclei in each voxel and of the imaging technique (the so-called MR sequence) used [34]. It has been shown about two decades ago that hyperpolarization techniques such as DNP and PHIP even allow imaging of $^{13}$C and $^{15}$N labeled endogenous substances and their metabolism due to their high initial non-equilibrium polarization [35, 36].

As a hybrid approach spatially resolved spectroscopic techniques such as single-voxel spectroscopy (*Point Resolved SpectroScopy, PRESS*) or *Chemical Shift Imaging (CSI)* with usually larger voxel sizes can be used. However, these time-consuming techniques are not part of the daily clinical routine, but mainly used in research, e.g. in the analysis of brain metabolites or muscle energy consumption [37, 38]. The experimental conditions in MRI are somewhat different from those in spectroscopy. Using sequence-dependent excitation pulse angles the echos may be formed by refocusing radio frequency (RF) pulses or by fast magnetic gradients as well as. Spatial resolution requires applying additional magnetic gradients in all spatial directions and adjustment of echo- and repetition times (TE, TR) for achieving the desired image contrast and clinical information [34].

The lower signal gains achieved with $^{19}$F photo-CIDNP are sufficient for MR imaging [39], as the cyclic response allows multiple excitations and thus signal averaging to increase SNR. However, the achievable resolution and optimal experimental conditions have not yet been investigated in more detail.

Photo-CIDNP offers several advantages that make this technique very suitable, especially for use in a medical environment: First, potentially hazardous gases such as hydrogen are not required, making it much safer to use. Second, polarization can be generated many times in a defined manner, which allows averaging the signal and increasing the SNR. Third, there are biocompatible solvents and molecules, so there is no need to remove toxic catalysts and solvents prior to application. Fourth, hyperpolarization generation requires very little and cheap hardware and is quite mobile. The replacement of lasers by high-power LEDs for photo-CIDNP studies was successfully investigated by [39–42] and references therein.

Due to much lower signal amplification (SE) compared to PHIP- or DNP-based methods, 1H photo-CIDNP in living organisms requires specific water and fat suppression techniques in 1H MRI. Otherwise, signal enhancement may not be detectable. However, in spectroscopic imaging, water and fat suppression is not mandatory. In both 1D and 2D MR techniques, valuable information can be obtained by changing the signal intensity or phase. Recently, advances have been made to detect even the smallest concentrations or amounts of substances [6, 42–44] but the focus in this study is more on investigating the feasibility of a background-free approach using $^{19}$F NMR and $^{19}$F MRI.

Despite the advantages of photo-CIDNP, there are only few reports on spatially-resolved photo-CIDNP [39, 45, 46] and photo-CIDNP MRI has not yet been systemically studied. An important factor is the field dependence of the CIDNP effect [23]. Low-field conditions facilitate the transfer of polarization between heteronuclei, while standard MR imaging and NMR spectroscopy generally profit from higher $B_0$ fields. Photo-CIDNP may therefore exhibit greater signal enhancement at low fields [23, 47]. Recently, photo-CIDNP has been reported in earth magnet field or even towards zero field strength [28]. However, the $B_0$ dependency varies strongly for different molecules and different nuclei [23, 24] which requires to investigate each model system separately. Theoretical calculations are here of great value [11, 23].

To optimize the polarization of heteronuclei, it would be of great advantage to disentangle the transmission of the generated polarization between different nuclei within the molecule of interest. However, at high fields, this can only be achieved with state-of-the-art technology due to the larger gap between the corresponding Larmor frequencies [48]. At very low and ultralow fields, broadband magnetic sensors such as SQUIDs can be used to measure different nuclei simultaneously because there is little difference in the Larmor frequencies [47, 49]. Recently, this technique enabled the detection of hyperpolarization based on parahydrogen (pH$_2$) and provided insights into the underlying quantum mechanical mechanisms [50]. However, SQUIDs are expensive and require thorough magnetic shielding from the environment, limiting their use to specially equipped laboratories. Here, we take advantage of the fact that at 0.6 T the Larmor frequencies of $^1$H and $^{19}$F are close enough to be simultaneously detectable with a conventional Faraday effect-based coil optimized for $^{19}$F.

Summarizing our goals, we aimed to investigate the feasibility of low-cost MRI at 0.6 T using hyperpolarized $^{19}$F in a biocompatible model system. To the best of our knowledge, there are no published data for $^{19}$F photo-CIDNP-based MRI of biocompatible molecules and solvents at medium-high fields. To determine the signal enhancement a comparison of spectra with and without illumination was most appropriate. However, the benchtop MRI used here is optimized for imaging, and therefore we first had to optimize spectroscopic data acquisition with a focus on multinuclear spectroscopy.

As the $B_0$ field of 0.6 T is significantly lower than the $B_0$ used by state-of-the-art clinical scanners (being in the range of 1.5 T to 3 T and more recently 7 T), we expected that non-hyperpolarized substances would be undetectable without considerable averaging. Although the magnet used in our experiments is temperature controlled, some residual fluctuations remained, resulting in small fluctuations in the signal frequency. It was therefore necessary to test whether a manufacturer-provided averaging algorithm with integrated correction of these fluctuations would enable a reliable long-term data acquisition and averaging. We also wanted to analyze to the extent to which imaging data can help to more accurately estimate the signal enhancement in spectroscopic measurements when uniform illumination is not achieved within the sample volume, as expected in biomedical applications.

The paper is organized as follows: First, we describe the optimization of the experimental setup for simultaneous multinuclear spectroscopic measurements, including spectral resolution and stability of long-term data acquisition. Then, representative results of hyperpolarized $^{19}$F imaging are presented. Finally, the photo-CIDNP based $^{19}$F signal enhancement at 0.6 T in liquid phase is estimated, taking into account the information derived from the images.

## 2 Materials and Methods

### 2.1 Sample preparation

A standard stock solution was prepared by dissolving 1.205 mg (6.025 µmol) 3-fluoro-DL-tyrosine ($C_9H_{10}FNO_3$, 199.18 g/mol, CAS 403-90-7, ABCR) and 0.4 mg (0.836 µmol) riboflavin 5′-monophosphate sodium salt hydrate ($C_{17}H_{20}N_4NaO_9P \cdot xH_2O$, 478.33 g/mol, CAS 130-40-5, Sigma) in 3 mL 99.9 % $D_2O$ (Deutero GmbH).

A volume of 580 µL of the stock solution containing only $D_2O$ was filled into a 10 mm NMR tube. Due to the low $B_0$ and the low concentration of 3-fluoro-D/L-tyrosine, the sample did not show a detectable $^1$H or $^{19}$F NMR signal in single scans that would have allowed reliable shimming. Therefore, we added 20 µL $H_2O_{dest.}$ (distilled water) and used the $^1$H channel for both shimming and $^1$H imaging to control the optimum position of the sample. The resulting amount of substance of 3-fluoro-D/L-tyrosine was 1.165 µmol with a concentration of 1.94 mmol/L, and the amount of riboflavin 5′-monophosphate sodium salt hydrate was equivalent to 0.156 µmol with a concentration of 0.27 mmol/L. The stock solution was stored in the dark at about 277 K and could usually be used for up to several days.

To calibrate the detection limit and coil sensitivity, we used pure $H_2O_{dest.}$ or 2,2,2-trifluoroethanol (TFE, $C_2H_3F_3O$) solutions in 10 mm and 5 mm glass tubes. To analyze experimental conditions such as spectral resolution or long-term data acquisition of weak signals (i.e. solutions with low substance concentrations) we used both 100% TFE and highly diluted TFE in $H_2O_{dest.}$. The latter solution exhibited a $^{19}$F concentration of 2.01 mM, which is approximately the $^{19}$F concentration of 3-fluoro-D/L-tyrosine in the later experiments.

## 2.2 Light excitation

A low-cost high-power blue LED (CREE XP-E, high-power LED 455 nm) was used for photoexcitation of FMN. The light was coupled into an optical fiber (THORLABS, Ø1000 µm Core Multimode Fiber, Low OH) with maximum intensity determined to be 900 µW in a previous experiment (further characterization of the LED and additional technical details are described in detail in the supporting information (SI) in [39]). A 3D-printed cylinder with a central hole of 1 mm was inserted in the glass tube without contact to the solution to center the fiber in the sample. The tip of the fiber was placed in the middle of the solution (Fig. 2).

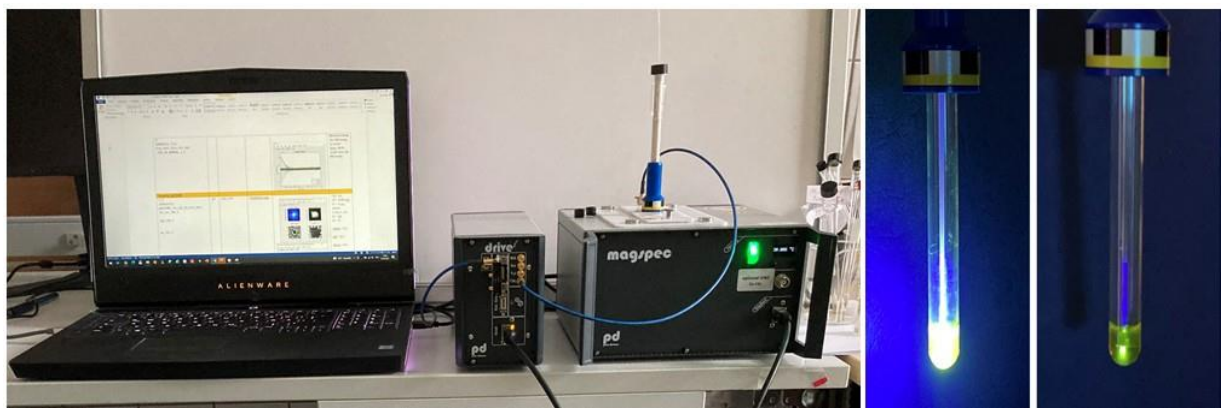

**Fig. 2** Experimental setup
*Left image:* MR benchtop system with laptop, research control unit and magnet with removable $^{19}$F coil.
*Middle image*: NMR tube containing 3-fluoro-D/L-tyrosine/riboflavin 5'-monophosphate. The glass fiber is centrally positioned, the tip of the glass fiber is located approximately in the middle of the sample volume. The sample is illuminated with high intensity light.
*Right image*: The stronger illumination below the glass fiber tip and the reduced illumination in the upper half of the sample volume where light exits laterally from the glass fiber is more clearly seen under low intensity conditions.

## 2.3 Data acquisition

Experiments were performed with a MR unit (Research Magnet magspec, Pure Devices GmbH, Würzburg, Germany) equipped with gradients of 250 mT/m (x-,y-direction) and 350 mT/m (z-direction) and a permanent magnet of about 0.6 T. The magnet was equipped with a replaceable $^{19}$F coil unit, and the optimal measurement volume was 1 cm$^3$ isotropic. The unit provided linear but no higher-order shims. Magnet, coil and gradients were connected to a control unit (Research Console Unit, Pure Devices GmbH, Würzburg Germany), which was connected to a standard laptop. The permanent magnet was temperature-stabilized at 303 K.

For imaging, we used the *version 2020* of the firmware *OpenMatlab* provided by the vendor, running on *Matlab 2020b* (Mathworks, Natick, MA, USA). Although *version 2020* of *OpenMatlab* also allows spectroscopy, we used an earlier version (*version 2017*, *OpenMatlab*) of the firmware because this version allowed to run an additional module for simultaneous multichannel spectroscopy. For long-term spectroscopic measurements, an additional 20 Ω resistor was inserted between the controller and the gradients to stabilize the shim.

At the beginning of the measurement, a sequence provided by the manufacturer was run to automatically determine and optimize the most important parameters (90° pulse duration, shim values, frequencies, etc.). If the $^{19}$F signal was very weak, but the $^1$H signal was sufficient, the shim was performed manually on the $^1$H channel and the remaining parameter values were taken from previously acquired values. Several custom *Matlab*-based programs were developed for manual phase correction of the signals, calculation of signal intensities by integrating the corresponding real parts of the Fourier-transformed spectrum, and determination of SNR.

In order to both capture images and determine the signal enhancement of the same solution, we developed a complex dual channel measurement protocol that alternates between imaging and spectroscopy. The latter required the insertion of a resistor (s. above) and therefore a new shim as well as using different versions of the control software. The overall process typically consisted therefore of the following steps: (a) initially checking the degree of hyperpolarization, (b) checking the position of the sample with $^1$H imaging, (c) capturing hyperpolarized $^{19}$F spectra, (c) capturing $^{19}$F images with and without illumination under different conditions and finally (d) recording of $^{19}$F reference spectra without illumination (see SI Fig. 1). The acquisition of the non-illuminated reference data for both imaging and spectroscopy required long-term measurements of several thousand runs. In order to minimize the risk of generating light-induced by-products, the total duration of illumination in steps a and c had to be small.

### 2.3.1 Imaging

A 2D turbo spin-echo (TSE) sequence was used for imaging. If not otherwise stated, we used a gauss-shaped pulse with a duration of 40 µs for the 90° excitation RF pulse, TR = 15 s, TE = 6 ms, a read- and phase-encoding oversampling factor of 4, the turbo factor set equal to the number of phase-encoded lines, a field of view 15 mm x 15 mm, and 1 slice covering the entire sample. The sequence averaged the acquired data during the measurement, and also stored the raw data and post-processed data of each single image for further processing. The software included a user interface to display zero-filled image data, k-space data and their respective phase images.

### 2.3.2 MR Spectroscopy

For dual-channel spectroscopy, a single rf send-and-receive coil tuned to the Larmor frequency of the $^{19}$F nuclei was used. For dual-channel spectroscopy, a single rf send-and-receive coil tuned to the Larmor frequency of the $^{19}$F nuclei was used. Therefore, the energy of the composite pulse had to be increased to cover both RF pulses. Since the coil efficiency at the frequency for the $^{1}$H nuclei was lower than the efficiency at the $^{19}$F frequency, the pulse energy was divided into two parts: 10 % was transmitted at the $^{19}$F channel (correlating approximately to a 90° pulse condition with typically 40 µs pulse duration), while 90 % was transmitted at the $^{1}$H channel (corresponding to a typical pulse duration of 360 µs). Larmor frequencies could be set independently for each channel according to the resonance lines of interest. For 3-fluoro-D/L-tyrosine this resulted in a typical chemical shift of about -1150 Hz relative to the main $^{19}$F spectral peak of TFE.

Long-term spectroscopic measurements, in which several thousand free induction decay (FID) signals were acquired, required post-processing of the acquired FIDs prior to averaging because small fluctuations in the $B_0$ field strength were still present. Post-processing included phase correction and the application of frequency filters prior to averaging the FIDs. The corresponding *Matlab* routine was provided by Pure Devices GmbH (it is not part of the standard software package). For TFE dissolved in $H_2O_{dest.}$, the Larmor frequencies were typically 24,341,219 Hz for $^{1}$H and 22,901,550 Hz for $^{19}$F.

### 2.3.3 Optimization of experimental conditions

Prior to the actual measurements, we determined the optimal conditions for spectroscopic data acquisition and tested the reliability of the long-term measurements, including subsequent post-processing with pure TFE. First, we used TFE to determine the sensitivity of the $^{19}$F coil when the transmit and receive frequencies were varied. Since the maximum resonance line of 3-fluoro-D/L-tyrosine is about -1150 Hz lower than that of TFE, a corresponding change of the frequency resulted in a decrease of about 20% in the signal (SI Fig. 2), which still allows reliable detection of 3-fluoro-D/L-tyrosine.

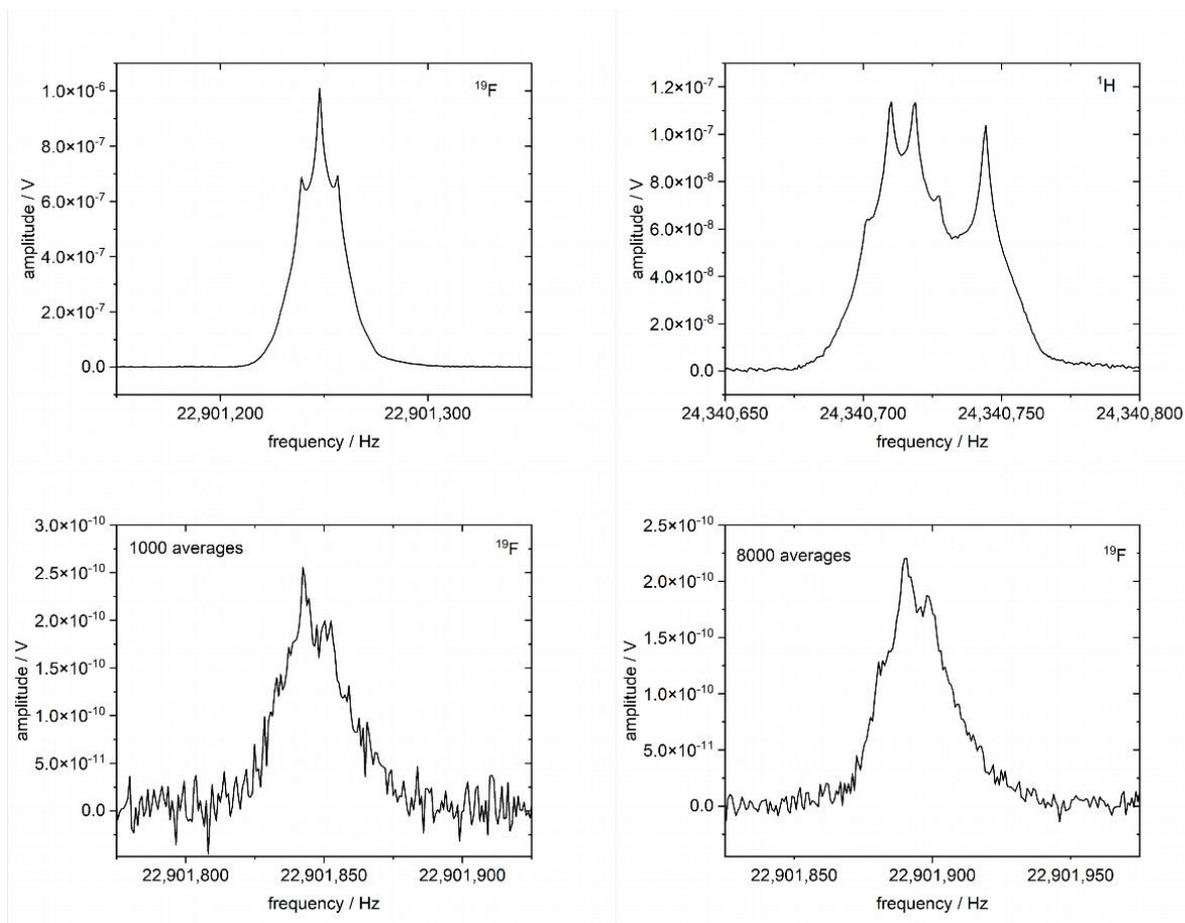

**Fig. 3** NMR spectra of pure TFE and highly diluted TFE in $H_2O_{dest.}$ at 0.6 T.
*Upper row*: Spectroscopy of pure TFE. Phase-corrected real parts of $^1H$ (left) and $^{19}F$ spectra (right) (sample volume 600 μL in 10 mm glass tube, bandwidth 128 kHz, 178k data points, 1 FID acquired). The $^1H$ spectrum shows the quartet of the $CH_2$ group at 24,340,710 Hz and a broad single resonance of the OH group at 24,340,750 Hz (please note that the frequency increases towards the right side). The $^{19}F$ spectrum depicts the $^{19}F$ triplet at 22,901,250 Hz.
*Lower panel*: Test of the stability of the experimental setup and of the quality of the correction algorithm. Eight thousand $^{19}F$ FIDs of diluted TFE in $H_2O_{dest.}$ were acquired and averaged (420 μL in 10 mm glass tube with a concentration of $^{19}F$ nuclei of 2.01 mmol/L which is approximately equivalent to the $^{19}F$ concentration of 1.94 mmol/L 3-fluoro-D/L-tyrosine; bandwidth 16 kHz, 22k data points). Averaging results in a mean frequency, which thus may be altered by the number of signals.

In the next step, we determined the spectral resolution for large sample volumes in a 10 mm glass tube, because smaller glass tubes would allow a better shim but would further reduce the signal of the 3-fluoro-D/L-tyrosine. $^1H$ and $^{19}F$ signals from pure TFE were detected simultaneously with the $^{19}F$ coil. The aim was to investigate whether the spectral resolution was sufficient (a) to distinguish the signals associated with the chemical shift within the corresponding $^1H$ and $^{19}F$ frequency ranges, (b) how the signals compared with standard high-field NMR, and (c) whether the sensitivity of the $^{19}F$ coil to $^1H$ would still be high enough to record good $^1H$ spectra. Fig. 3 shows that the spectra for both $^1H$ and $^{19}F$ signals have good resolution and signal strength. The integrals of both spectra were found to be $2.37 \cdot 10^{-5}$ V·Hz for $^{19}F$ and $4.35 \cdot 10^{-6}$ V·Hz for $^1H$. Since TFE has the same number of $^1H$ and $^{19}F$ nuclei this corresponds to a sensitivity of the $^{19}F$ coil for $^1H$ of about 17 % under the experimental conditions (taking into account the 6% smaller gyromagnetic ratio of $^{19}F$). An additional assumption was that no further protons contribute to the signal.

The $^1H$ and $^{19}F$ spectrum allowed the detection of the typical spectral patterns of TFE known from high-field NMR. The resolution was good enough to determine the $^3J(H,F)$ coupling constant to be about 9 Hz. This agrees well with previously reported values of the $^3J(H,F)$-coupling [49]. Compared to high-field NMR spectra, the signals are broadened on the base, likely due to residual field inhomogeneity within the relatively large sample volume of 1 $cm^3$ and the lack of non-linear shim coils. This could complicate detailed analysis of other substances that have more complex spectra or spectra with lower amplitude components.

As expected, non-hyperpolarized 3-fluoro-D/L-tyrosine was not detectable in single measurements even after optimizing the experimental settings. To test how many averages would be necessary to reliably detect a signal with the approximate concentration of the 3-fluoro-D/L-tyrosine, and whether subsequent averaging of the corrected FIDs would lead to reliable results, highly diluted TFE in $H_2O_{dest.}$ (2.01 mmol/L, approx. equivalent to the $^{19}F$ concentration of 1.94 mmol/L 3-fluoro-D/L-tyrosine) was measured in a 10 mm glass tube (420 μL sample volume, 16 kHz bandwidth, 22k data points, measurement duration 61.3 h). After about thousand averages the $^{19}F$ triplet began to be recognizable, but several thousands more measurements were needed for a clear characterization (Fig. 3, lower panel). The results also clearly show that the small remaining fluctuations in the signal, due to the temperature fluctuations of the magnet, were well corrected by the averaging algorithm.

## 3 Results

### 3.1 Imaging

Figure 4 shows representative results of a measurement cycle in which images as well as spectra of illuminated and non-illuminated samples (including 3-fluoro-DL-tyrosine) were acquired for both $^1H$ and $^{19}F$. Each measurement cycle was started by acquiring $^1H$ images in two orthogonal orientations to check the position of the sample and the optical fiber (Fig. 4A-4D), since the $^{19}F$ MR signal is either too weak (non-illuminated sample) or would only show the illuminated part of the sample. To increase the signal strength, the slice thickness encompasses the entire sample volume, resulting in an effective projection of the sample onto the corresponding image plane. Therefore, the void due to the optical fiber is only visible in the axial orientation, as the fiber is running along approximately half of the vertical axis within the sample, while in the sagittal orientation the projection of the sample volume obscures the according signal void.

Figs. 4E-4H show that up to several thousand images must be averaged to image non-illuminated 3-fluoro-D/L-tyrosine. Although initial details of the sample were apparent after a few hundred averages, it is only after about thousand averages that the image begins to show the structure of the object in more detail, including the central signal void caused by the optical fiber. After two thousand images, the structure of the object was represented with an acceptable quality. The repetition time was set to 15 s to allow full recovery of the longitudinal magnetization. This resulted in a total measurement time of about 500 min or 8.33 hours. The experimental setup proved to be stable over this period. Figures 4I-4K show the dramatic shortening in data acquisition time due to light-induced hyperpolarization. These images were acquired in one single acquisition lasting 11.3 s (including illumination time). Different matrix sizes were acquired to test the impact of resolution and zero-filling. Hyperpolarization is already clearly visible in the low-resolution images (Figs. 4I, 4K). However, higher resolution is required for smaller structures. Although hyperpolarization was detectable in individual higher resolution images, the lower SNR required additional averaging. Therefore, eight images were acquired under continuous illumination. Each image showed hyperpolarization, and even the eighth image (Fig. 4O) did not exhibit significant signal loss. Averaging resulted in better image quality (Fig. 4P). Additionally, zero-filling, an implicit interpolation method, allowed us to distinguish more clearly several sub-regions with different levels of hyperpolarization, which is a pre-requisite for determining the spatial distribution of the signal enhancement (see next section): The upper, less hyperintense part corresponded to the parts of the solution illuminated by the lower intensity light emerging laterally from the optical fiber while the lower, more hyperintense part corresponded to the regions of the solution directly illuminated by the high intensity light emerging vertically from the fiber. Note that the frequency-encoded spatial localization between $^1H$ and $^{19}F$ cannot be directly compared due to the different Larmor frequencies of $^1H$ and $^{19}F$. However, the non-illuminated $^{19}F$ MR image can be used as a reference, and it can be seen that the position of the axially located hyperpolarized spot coincides well with the dark gray spot on the axial non-illuminated $^{19}F$ image.

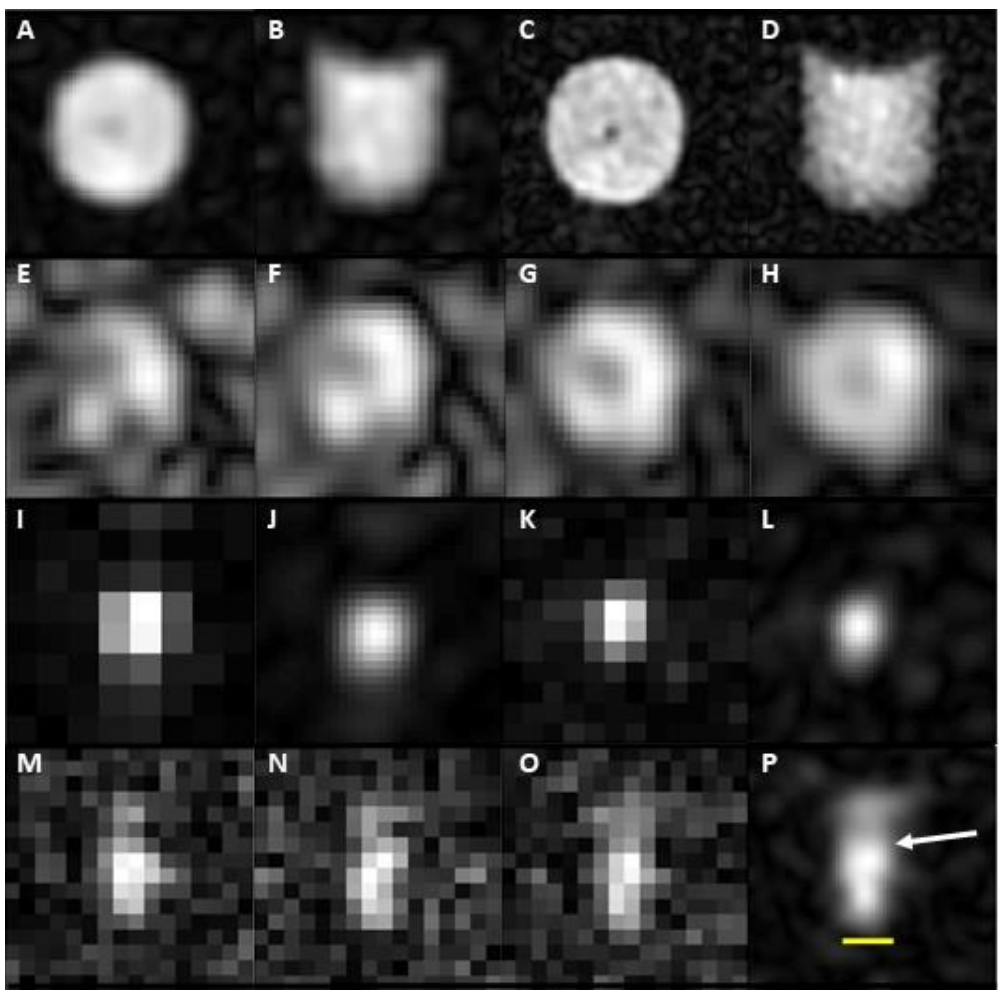

**Fig. 4** Representative TSE images of $^1$H and $^{19}$F MR imaging of a non-illuminated and an illuminated sample of a D$_2$O+H$_2$O$_{dest.}$ solution containing 3-fluoro-D/L-tyrosine and riboflavin 5'-monophosphate. Signal intensities were scaled to achieve the highest contrast.

**A-D:** $^1$H MR imaging without illumination to check position of the sample and optical fiber.

Axial (A, orthogonal to the vertical axis of the sample) and sagittal (B, parallel to the vertical axis of the sample) orientation, 1 average, matrix size 16 x 16, zero-filling factor 4. Axial (C) and sagittal (D) orientation, 8 averages, matrix size 32 x 32, zero-filling factor 4.

The central signal decrease in the axial images shows the projection of the optical fiber along the sample.

**E-H:** $^{19}$F MRI of a non-illuminated sample of 3-fluoro-D/L-tyrosine as a function of the number of averages.

A total of 2000 averages were acquired (matrix 8 x 8 voxel, axial orientation). The central void is due to the optical fiber. Data were zero-filled (factor 4) resulting in smoothing of both object and noise data. From left to right: 400 averages (E), 650 averages (F), 1000 averages (G), 2000 averages (H).

**I-L:** $^{19}$F MR images (axial orientation) of the illuminated sample of 3-fluoro-D/L-tyrosine.

I: Matrix size 8 x 8 voxel, no zero-filling, original resolution, one average. The hyperintense voxel represents the projection of the hyperpolarized volume along the vertical axis of the sample.

J: Same image as in Fig. 4I, but with zero filling factor 4.

K: Increasing the matrix size to 12 x 12 voxel, no zero filling.

L: Same image as in Fig. 4K, but with zero filling factor 4.

**M-P:** Representative high-resolution $^{19}$F images of the illuminated sample of 3-fluoro-D/L-tyrosine (sagittal orientation) and the resulting averaged image (eight averages, zero filling factor 4, sagittal orientation).

M-O: Three representative single images of an averaging cycle of eight images (16 x 16 voxel, no zero filling). Shown are the first, the third and the eighth image.

P: Averaged image (eight images, zero filling factor 4). The white arrow shows the position of the optical fiber tip (entering the sample from the above), the yellow bar represents the approximate diameter of the hyperpolarized volume (about 2.8 mm). Below the optical fiber tip, the signal intensity reaches a maximum value.

## 3.2 Estimation of $^{19}$F MR signal enhancement factor

Initially, we could not detect hyperpolarization in the spectra, although hyperpolarized signals were clearly seen in the imaging experiments, confirming that hyperpolarization was detectable at 0.6 T. However, to estimate the signal enhancement factor more precisely and compare it with previous results from spectroscopic experiments, it was indispensable to compare spectroscopic measurements of non-illuminated vs. illuminated signals. To obtain an optimal $^{19}$F hyperpolarized signal, several factors were crucial: First, the transmission pulses for 3-fluoro-D/L-tyrosine had to be fairly accurately on-resonant, otherwise the oscillations in the FID and the reduced excitation and detection amplitudes due to the frequency shift would make a clear detection of the signal difficult. Second, the acquisition time had to be sufficiently short, since the $T_2^*$ time of the hyperpolarized sample was found to be around 30 ms even under optimum shim conditions (SI Fig. 3 and SI Fig. 4). Long acquisition times, as used for TFE, unacceptably increased the noise level and thus hindered the detection of the signal. In summary, the optimum protocol for an acceptable spectral resolution with minimal noise is between 1000 and 4000 data points with a bandwidth between 16 kHz and 32 kHz, since the benchtop MRI exhibited a minimum bandwidth of about 16 kHz. In general, we used 22 kHz bandwidth and 2000 data points.

Our goal was to perform imaging and spectroscopy on both channels using the same sample if possible. However, since the $^{19}$F nucleus had only a very weak signal, we added additional $H_2O_{dest.}$ for the different calibration steps and measurements (SI Fig. 1). The reason for this was to reduce the possible formation of by-products in the sample, caused either by prolonged irradiation or by side reactions during long examination times when the sample remained in the magnet at 303 K. For long-term spectroscopy, a resistor must be inserted to reduce fluctuations of the shim gradients, while imaging must be performed without a resistor. Changing the resistor required a new shim each time. If only spectroscopy or imaging is performed, the protocol can be simplified accordingly. We limited the total irradiation in checking the hyperpolarization level to a few spectroscopic and imaging measurements with only minimal averaging before the actual measurements with illumination (SI Fig. 1).

We started to determine the signal enhancement by applying the standard procedure of comparing spectra with and without illumination. We used a new sample from the same stock volume (kept in the refrigerator at around 277 K for 2 days) to avoid measuring by-products due to the two-day measurement time of the reference non-illuminated image. Spectra were acquired with 16 kHz bandwidth (the minimum) and 4k sample points (other experimental conditions remained unchanged leading to an illumination time of 36 s and an overall measurement time for one spectrum of 41 s. The longer time of the measurement relative to the shorter acquisition time of the spectrum resulted from inclusion of additional calibrations such as adapting the frequency). After acquiring and averaging eight hyperpolarized FIDs non-illuminated signals were acquired for reference. In order to achieve sufficient SNR we had to average 8000 FIDs (Fig. 5) leading to an overall data acquisition time of 91 h (3.8 days).

However, when determining the signal enhancement factor, it must be taken into account that only a small part of the sample volume was hyperpolarized. Compared to the signal of the non-illuminated sample, which originates from the entire sample, a volume correction factor must therefore be applied when determining the signal enhancement: The hyperintense region in Fig. 4P can be divided, as a first approximation, into a lower part with higher signal intensity and an upper part with lower signal intensity (the white arrow represents the boundary between the two parts).

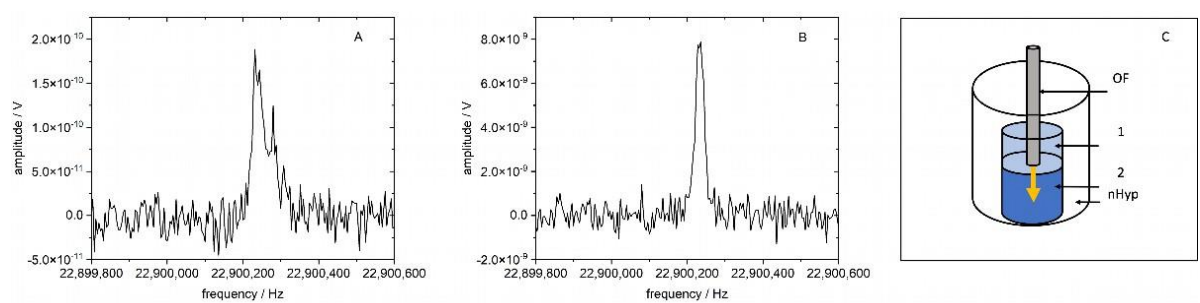

**Fig. 5** $^{19}$F spectra of 3-fluoro-D/L-tyrosine/riboflavin 5'-monophosphate in $D_2O+H_2O_{dest.}$ solution.
*A*: Non-illuminated sample. The curve shows the phase-corrected real part of the Fourier-transformed average of eight thousand FIDs acquired with a bandwidth of 16 kHz and a sampling rate of 4000 data points.
*B*: The same sample under illumination (average of eight FIDs, other conditions the same as in *A*). In each measurement the sample was illuminated 15 s prior to data acquisition and during each measurement cycle resulting in a total illumination time of 36 s. Dark time between measurements was 5 s resulting in a total measurement time of 41 s for one single FID.
*C*: Schematic view of the NMR tube with the optical fiber (OF), the two partial volumes of the solution illuminated with low (*1*) and high (*2*) light intensity, and the part, where illumination is too weak to generate detectable hyperpolarization (*nHyp*).

We assumed that the intensity was proportional, to a first approximation, to the degree of hyperpolarization of 3-fluoro-D/L-tyrosine and that the intensity of irradiation is low enough not to cause significant bleaching or saturation effects. The high-intensity part (part 2 in Fig. 5C) was estimated to be about 5.2 mm long and 2.8 mm in diameter (neglecting the minor diameter at the bottom of the glass tube). The low-intensity part contained the optical fiber (part 1 in Fig. 5C). The image shows that hyperpolarization in part 1 was detectable to extend about 0.9 mm into the solution on both sides, assuming a radius of the optical fiber of 0.5 mm and an overall lateral extension of 2.8 mm. The length of part 1 was estimated to about 3.3 mm.

The volumes and intensities were estimated from the zero filled image Fig. 4P. The volume $V_1$ of part 1 was approximated by a ring of length $l_1$ = 3.3 mm, outer radius $r_{out}$ of 1.4 mm and inner radius $r_{in}$ of 0.5 mm (corresponding to the diameter of the optical fiber of 1 mm). The mean intensity of all voxels in part 1 was about 60% of the voxels in the high intensity part 2, which we interpreted to mean that 60% fewer molecules were hyperpolarized. The volume $V_2$ of the high-intensity part 2 was approximated by a cylinder with length $l_2$ = 5.2 mm and radius $r_{cyl}$ = 1.4 mm.

Assuming that the number of hyperpolarized molecules is proportional to the amplitude of the signal and the volume of the hyperintense regions, the number of hyperpolarized molecules $N_{hyp}$ was estimated (with c denoting an unknown proportionality constant) to

(1)     $N_{hyp} = c \cdot (0.6 \cdot V_1 + V_2) = c \cdot 42.6 \text{ mm}^3 = c \cdot 42.6 \text{ μL}$.

To determine the signal enhancement factor, consider that the reference signal reflects the signal of all molecules in the entire non-illuminated sample, which is proportional to a volume of

(2)     $N_{total} = c \cdot V_{total} = c \cdot 600 \text{ μL}$.

The reference signal of the non-illuminated sample must be scaled by a correction factor

(3)     $F_{CORR} = N_{hyp} / N_{total} = (0.6 \cdot V_1 + V_2) / V_{total} = 42.6/600 = 0.07$.

The $^{19}$F signal strength with ($I_{LED\_ON}$) and without illumination ($I_{LED\_OFF}$) was determined by integrating the phase-corrected real part of the averaged spectra (Fig. 5) over a range of 470 Hz (where both signal approached the baseline). This led to a signal $I_{LED\_ON} = 2.68 \cdot 10^{-7}$ V·Hz and $I_{LED\_OFF} = 8.23 \cdot 10^{-9}$ V·Hz. Taking into account the correction factor, the signal-enhancement was estimated to

(4)     $SE = I_{LED\_ON} / (I_{LED\_OFF} \cdot F_{CORR}) = 268/(8.23 \cdot 0.07) = 465$.

## 4 Discussion

Tyrosine is a promising biomarker both for imaging and detection of metabolic processes as tyrosine is known to be a precursor of important neurotransmitters and tyrosine-kinases play an important role in signal transduction including tumor metabolism [51]. Recently, cerebral O-(2-$^{18}$F-fluoroethyl)-L-tyrosine uptake in rat epilepsy models was reported by combining positron emission tomography and MRI [52]. In further studies, the tyrosine concentration was associated with phenylketonuria. An example is the article by Waisbren et al. [53].

The results show that the goals of the project were achieved. The LED-based photo-CIDNP for the $^{19}$F biomarker in the 3-fluoro-D/L-tyrosine/FMN solution resulted in a strong signal increase, which compensated well for the lower SNR caused by the lower field strength of 0.6 T and the lower light energy of the LED. In addition, the signal amplification significantly shortened the total measurement time, since only a few images had to be acquired depending on the resolution. Sub-mm high-resolution $^{19}$F imaging was successful even with a few averages. The experimental setup, including post-processing algorithms, also enabled reliable long-term measurements with acceptable spectral resolution. As a final plus, the two-channel NMR allowed for the simultaneous acquisition of $^1$H and $^{19}$F spectra with nearly comparable quality using the $^{19}$F coil. The various aspects are discussed in more detail below.

*Experimental setup*

With a total cost of well below 100,000 €, the benchtop MRI scanner including two coils is well affordable, takes up very little space and is easy to transport. While the experimental setup is primarily designed and optimized for imaging, spectroscopic measurements (including multi-channel spectroscopy and data post-processing) could also be performed in acceptable quality despite the lower field strength. Weak signals requiring considerable averaging and measurements over several days can be averaged with excellent quality. The *Matlab*-based control and processing software is open to the user and allows for customization and proprietary enhancements such as digital filtering. The combination of spectroscopy with imaging was also a major advantage, allowing reliable verification of the position of the probe and optical fiber position. This function is usually only available in clinical scanners or high-field spectrometers with additional costly gradient inserts. Another major advantage for our purposes was the attractive possibility of simultaneously acquiring NMR signals from several nuclei with only one transmission and detection unit, since the $^{19}$F coil was sufficiently sensitive at the $^1$H frequency.

Since the vast majority of published data on the photo-CIDNP effect are based on spectroscopic experiments, we decided to include spectroscopic experiments in our study to allow comparison with other results. Using TFE as a standard to characterize the spectral resolution we were able to unambiguously determine the $^3$J(H,F) coupling to 9 Hz, confirming earlier results [39]. However, the non-Lorentzian broadened line shape will affect the resolution of spectra with more complex spectral patterns. A smaller volume could increase the shim quality and thus spectral resolution, of spectra but this will result to lower SNR or prolonged measurement time. There are also more sophisticated post-processing algorithms to reduce noise [54] or to use adapted filtering techniques.

The accumulation of several thousand signals took several days, but proved to be stable. Compared to our previous high-field measurements [39] with a repetition time of 60 s, we chose a repetition time of 15 s as a compromise between measurement duration and signal strength. In future measurements, we will analyze the decay of the hyperpolarization and the $T_1$ relaxation time of 3-fluoro-D/L-tyrosine at 0.6 T in more detail to adjust the experimental conditions optimally. Another important factor could be the influence of bleaching of the flavin, which, however, seemed to be less pronounced in LED-induced photo-CIDNP in the 10 mm glass tube than when lasers were used in 5 mm glass tubes. Although we did not yet perform a detailed analysis, we found that hyperpolarized signals could be detected up to about 50 times without significant bleaching. The sample could be hyperpolarized up to about one week after initial use when stored in the dark at 277 K.

Although the multinuclear capability should allow direct observation of simultaneous polarization of different nuclei, we did not detect significant hyperpolarization of the $^1$H nuclei in 3-fluoro-D/L-tyrosine despite previous reports and theoretical calculations [11, 39]. Most likely, the hyperpolarization level is still too weak compared to $^{19}$F and is further masked by the $^1$H MR signal from $H_2O_{dest.}$ added to the solution to enable shimming. However, similar conditions will occur when 3-fluoro-D/L-tyrosine is used as a molecular marker in a biological environment such as a cell, highlighting the importance of using a hyperpolarizable background-free marker molecule such as fluorinated substances.

*Imaging*

While hyperpolarization methods such as DNP, PHIP and their modifications have long been used for research purposes in NMR spectroscopy, it was their application for imaging and spatially resolved spectroscopic detection of certain metabolites that attracted the interest of medical professionals [35, 36]. We found that photo-CIDNP offered the optimal conditions, because not only was it the only method to offer biocompatible model systems, but also the overall handling was simple and safe. As we have shown in Fig. 4, the lower level of hyperpolarization, especially when increasing the spatial resolution, can be well compensated with the standard NMR averaging procedure. Averaging is limited only by chromophore bleaching or unwanted or potentially toxic by-products, which will be investigated in future studies.

Despite these encouraging results, few other publications have demonstrated the feasibility of spatially resolved photo-CIDNP. Trease et al. [45] obtained spatially-resolved spectral information of the hyperpolarized n-acetyltryptophan dissolved in pure $D_2O$ with 2,2-dipyridil as chromophore. Hyperpolarization was generated by a UV excimer laser (XeCl, 308 nm), and a spatially resolved chemical shift spectrum of the hyperpolarized sample was recorded at 300 MHz. The focus of the study was to demonstrate that spatial encoding can be significantly accelerated by CIDNP-mediated optical encoding instead of the usual encoding employing magnetic gradients. Spatially resolved $^{19}$F photo-CIDNP was also reported by Gueden-Silber et al. [46]. This group used an air-cooled laser diode (10 W) to hyperpolarize 7 mM 3-fluoro-DL-tyrosine/0.2 mM FMN in $D_2O$ and acquired spatially resolved data using a multi chemical shift-selective imaging (mCSSI) sequence.

Although we had published preliminary results of a spatially resolved hyperpolarized $^{19}$F image using a 2D TSE technique [39] the impact of varying the resolution and the number of averages have not yet been analyzed in more detail. Similar to spectroscopy, we also wanted to acquire a non-illuminated reference $^{19}$F image, primarily to determine the position of the probe and the optical fiber with respect to the hyperpolarized $^{19}$F image, since the $^1$H image could not readily be used to overlay a $^{19}$F image. Interestingly, the hyperintense signal was initially much clearer detectable in MR imaging than in the spectroscopic signal. There may be several additional factors that may contribute to this finding: the bandwidth in an image is smaller than in spectroscopy (here, we used 500 Hz per voxel in image data as compared to 22 kHz for one FID in the spectroscopic data).

In addition, short data acquisition times reduce the noise, but at the cost of lacking spectral resolution. In imaging, however, the acquisition of multiple spectral resonances within one voxel is usually undesirable because they lead to artifacts (e.g., the chemical shift artefact between fat and water). Moreover, the acquisition of one image already implies an inherent averaging of multiple hyperpolarized signals, as each phase-encoded row of the image matrix corresponds to a digitized spin-echo. We will analyze the corresponding effects more systematically in future studies, in particular how the hyperpolarization behaves under continuous vs. fractionated illumination, number of successive 90° excitation pulses, and varying echo times. As a first result, we found that at low spatial resolution (i.e., few echo trains) the acquisition of one image is already sufficient to localize the hyperpolarized region with good accuracy (Fig. 4I).

The image can be refined further by applying zero-filling (Fig. 4) which leads to interpolate the image information between adjacent voxels (zero filling is also often used in standard radiological diagnostics). As expected, increasing the actual resolution decreased the SNR and required averaging multiple hyperpolarized images. When bleaching does not result in significant signal reduction, as in our example, the fast TSE technique provides reliable images and can be used to acquire multiple images (Fig. 4 L-O). Kuprov et al. found, that when using a high power laser for excitation, he could irradiate 3-fluorotyrosine for more than 5 min [11]. Since our light source is much less intense than a laser, we expect to be able to average many more data sets than just the eight presented here, increasing the resolution much more.

*Signal enhancement*

Spectroscopic CIDNP studies of fluorinated molecules were reported shortly after the discovery of CIDNP by Bargon, Fischer, Ward and Lawler [7, 8, 55], but the substances studied were usually not bio-compatible. As a modification, photo-CIDNP developed as an interesting and important technique for the study of molecular reactions involving radicals and electron transfer processes. The effect was found to depend critically on the field strength of the external magnetic field and other factors. This field dependence was studied in great detail by the working group of Vieth [23, 24] over a wide range from near 0 T to 7 T using $^1$H photo-CIDNP for different non-fluorinated molecules including amino acids, chromophores and solutions. For tyrosine they found a maximum near 2 T (Fig. 5.21 in [24]) while the signals at 0.5 T and 7 T were comparable. A direct comparison with our results remains difficult because their experimental setup was different from ours. In addition to using other chromophores, solvents and excitation sources, they used a field-cycling technique to build up the hyperpolarization at different positions outside the spectrometer, while the detection was realized inside the spectrometer. Closer to the spectroscopic part of our study are the studies by the group of Kuprov, Hore et al. who investigated the $^1$H and $^{19}$F photo-CIDNP effect in fluorinated amino acids and proteins [10, 11, 16, 17]. The studies included a theoretical analysis taking into account hyperpolarization, cross-correlation and relaxation effects, as well as optimizing of sample illumination. Hyperpolarization was generated inside the spectrometer at 14.1 T using an argon laser.

Depending on the experimental conditions, they found up to 40-fold enhancement of the $^{19}$F magnetization. Comparing the model calculations with their findings, Kuprov et al. discussed the different interaction mechanisms of longitudinal relaxation, cross-relaxation, and cross-correlation with respect to the anisotropic hyperfine interaction and other parameters [11]. Unfortunately, the spectral resolution of our measurement of 3-fluoro-D/L-tyrosine does not allow us, comparable to our previous experiment [39], to analyze in more detail both the illuminated and non-illuminated $^{19}$F spectrum of 3-fluoro-D/L-tyrosine at 0.6 T in terms of changes in individual spectral lines. However, from Fig. 5, it may be speculated that hyperpolarization occurs only on selected resonances of the spectrum, since the half-width of the spectrum appears narrower under illumination than without illumination. This would be in line with our results at high-field, where only some of the spectral lines exhibited signal enhancement [39]. We will investigate this in more detail in future experiments.

There is also strong evidence that the signal enhancement for $^{19}$F in our model system is higher at 0.6 T than at 7 T. Interestingly, we did not detect any significant $^1$H signal enhancement even though the experimental setup enabled the simultaneous detection of $^1$H and $^{19}$F. It is likely, that the gain of the $^1$H signal is too low to overcome the thermic polarization of the $H_2O_{dest.}$ protons added to the solution to enable the shim procedure. In addition, rapid transfer (< 0.5 ms) of the $^1$H hyperpolarization to the $^{19}$F could prevent significant buildup of detectable $^1$H CIDNP, as data acquisition started approximately 100 μs after the $^1$H excitation pulse, with a duration of 360 μs. Further analysis of these rapidly occurring primary radical pair reactions [11] may be significantly improved by novel techniques such as combining NMR with ps- and ns-time domain detection using optical spectroscopy [56].

When discussing the $^{19}$F signal enhancement, it is important to keep in mind that the sample in our experiment was not illuminated homogeneously, whereas in most spectroscopic studies considerable effort has been made to illuminate the entire sample uniformly and at high intensity [28]. However, without spatially resolved intensity measurements, it is difficult to prove whether this goal was achieved. Imaging can provide excellent information on where and to what extent hyperpolarization is being produced. Taking this information into account and correcting the signal enhancement derived from the spectral measurements with a spatially resolved and intensity-weighted correction factor resulted in a signal amplification that was about 15 times higher than when simply comparing the spectral signals. However, the result may be preliminary as even small changes in the estimation of the degree of hyperpolarization and the hyperpolarized volume lead to significant changes of the signal enhancement factor. This result also cannot be readily compared to our earlier, much lower signal gain factor at 7 T, since we did not correct for possible inhomogeneous illumination of the sample there. In addition to the higher

$B_0$ field, the experimental conditions were not completely identical, so further measurements will be performed near future to more accurately determine the photo-CIDNP at 7 T.

We also found that it was difficult to derive the signal enhancement directly from the images, because the intensity varied greatly in the voxels and the partial volume effects were quite pronounced. When averaging over several voxels, we found a signal enhancement of only about 200 even for the most hyperintense voxels. The reasons for the discrepancy to the volume-corrected spectral analysis are not entirely clear. To rule out experimental causes, we compared the maximum signal in k-space (i.e. the maximum amplitude of the corresponding spin echoes) with the maximum signal of the averaged FID and found a very good agreement. Therefore, we suspect that the complex processes of acquiring raw data and converting them to image data may have an effect on the signal. However, further experiments are needed to investigate how noise, bandwidth, relaxation processes [27], spatial coding and other image sequence parameters as well as the decay of the hyperpolarization, will affect the contrast in the final Fourier-transformed image.

Finally, we estimated the amount of substance that can be detected in the hyperpolarized images. With a true in-plane resolution of $(0.9375 \text{ mm})^2$ and a sample thickness of 10 mm, a volume of 8.8 mm$^3$ can be resolved. This volume contained an amount of $17.1 \cdot 10^{-9}$ mol of 3-fluoro-D/L-tyrosine. Hyperpolarization thus made it possible to detect approx. 17 nmol of $^{19}$F within a few seconds, representing a major step toward increasing the speed and sensitivity of NMR for molecular imaging. Special experimental setups with microcoils have already led to the spectroscopic detection of very low concentrations or even sub-pmol of substance [41, 44]. The combination of these approaches may lead to a further increase in NMR sensitivity. In addition, simultaneous multinuclear detection is expected to lead to new insights into the transfer of hyperpolarization between different nuclei, even if only nmol of amounts of substance are present.

*Limitations of the study and outlook*

The study has some limitations as we initially focused on demonstrating the feasibility of hyperpolarized $^{19}$F imaging and determining the signal enhancement using spectroscopy. We did not determine the maximum signal enhancement as a function of the total irradiation time or the duration of the optimal dark times between successive measurements to reduce potential chromophore bleaching. To understand the optimized imaging conditions and determine the limiting number when averaging hyperpolarized signals, we still need to determine the decay of the hyperpolarization and the behavior of the signal under continuous vs. fractional illumination. To more accurately determine the spatial distribution and degree of hyperpolarization, we also need to increase the resolution and use three-dimensional MRI, as well as test other sequences such as gradient echo sequences and spatially resolved spectroscopic sequences. We will also use newer LEDs with higher output power to increase the efficiency of hyperpolarization. Finally, it should be noted that two factors at this stage of the study may limit biomedical applications. First, the prolonged high intensity irradiation can potentially produce phototoxic by-products [57]. Second, our solution consisted mainly of $D_2O$ (with 3.4% $H_2O_{dest.}$) and therefore may still be not be applicable to humans without some modification as $D_2O$ may exhibit some unwanted effects [58].

## 5 Conclusion

This study proves that the LED-induced photo-CIDNP effect enables reliable NMR spectroscopy and MR imaging of a $^{19}$F biomarker even at very weak signals and very low substance levels. It is well known that the photo-CIDNP effect is field dependent, and our results indicate that the signal enhancement is higher at 0.6 T than at 7 T. The use of hyperpolarization not only compensated for the significantly lower SNR at 0.6 T, but also allowed for a dramatic reduction in data acquisition time from several days to tens of seconds. In addition, multichannel NMR was shown to be of great advantage. In addition to detecting potential transfer processes between nuclei, the presence of a unique signal from another nucleus allowed the necessary calibration steps to perform reliable long-term acquisitions of other nuclei at very low concentrations. Even without fully optimizing the imaging procedures we achieved a sub-mm resolution in hyperpolarized images where one voxel exhibited about 17 nmol amount of substance. The information derived from the image about the spatial extent of the hyperpolarization and its location-dependent degree of enhancement not only allowed a more precise determination of the signal enhancement than using spectroscopic data only but can also indirectly serve as a measure for the spatially varying transformation of light into nuclear magnetization. Thus, the study shows how imaging and spectroscopy can benefit from each another. We expect that further improvements in imaging and hyperpolarization will pave the way to explore further new and exciting biocompatible applications. This might be very interesting for other biologically important molecules such as $^{13}$C, $^{15}$N, $^{35}$Cl, or $^{31}$P. Using the benchtop experimental setup offers an interesting option for the investigation of specific light-induced reactions and their applications. It may provide a complementary experimental setup to standard high-field NMR spectrometers and MRI scanners, especially for basic preclinical experiments.

## Data Availability

The data is available under request from the corresponding author.

## Statements and Declarations

The authors have nothing to declare.

# Supplementary Information

## LED-based Photo-CIDNP Hyperpolarization Enables $^{19}$F MR Imaging and $^{19}$F NMR Spectroscopy of 3-fluoro-DL-tyrosine at 0.6 T

Johannes Bernarding[1*], Christian Bruns[1], Isabell Prediger[1], Markus Plaumann[1]

**Affiliations**

[1*]Institute for Biometry and Medical Informatics, experimental medical imaging, Medical Faculty of the Otto-von-Guericke University Magdeburg, D-39120 Magdeburg, Leipziger Strasse 44

[*]Corresponding author: Johannes Bernarding, johannes.bernarding@med.ovgu.de

**Table of Content**

# 1 Measurement protocol for simultaneous dual channel spectroscopy and $^{19}F/^{1}H$ imaging for $^{19}F$ photo-CIDNP at 0.6 T

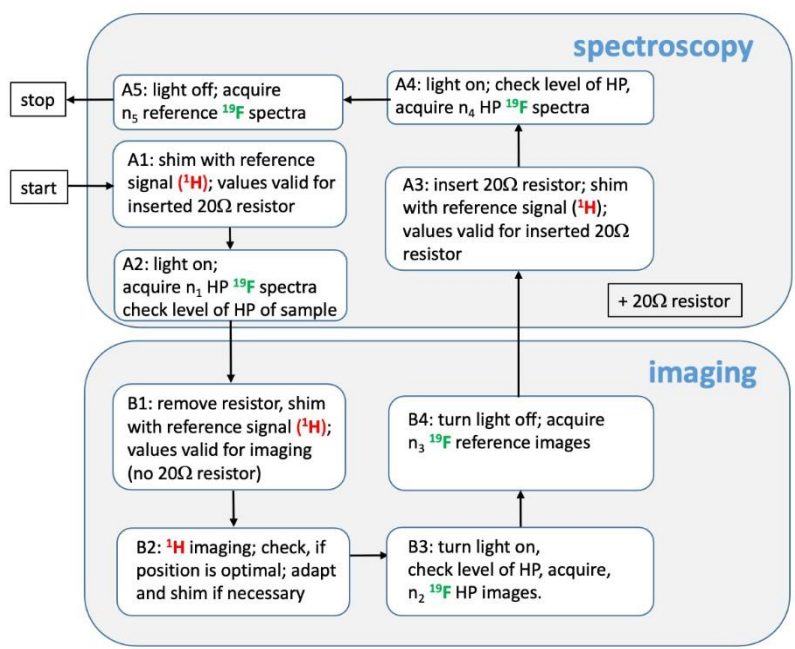

**SI Fig. 1**
Representative scheme of an optimized protocol for acquisition of spectroscopic and imaging data from the same sample (irradiated and non-irradiated). The scheme can be adapted if only one modality of data (spectroscopic or imaging data) is acquired.
$n_1$, $n_2$, and $n_4$ should be significantly smaller than $n_3$ and $n_5$ (number of non-irradiated reference measurements) in order to avoid irradiation-induced degradation of the sample.

A typical measurement cycle was composed as follows (the numbers of $n_i$ correspond to the measurements cycles, selected data of which are presented in the main text). *A* refers to spectroscopic measurements, *B* to imaging measurements.

a) acquire $^{19}F$ FIDs to check hyperpolarization level and Larmor frequency of sample ($n_1 = 9$);
b) acquire $^{1}H$ images in orthogonal directions to check localization of sample;
c) acquire hyperpolarized $^{19}F$ images (here, $n_2 = 15$ images were recorded with different resolutions and variations of other parameters; part of these images are presented in Fig. 4 of the manuscript);
d) acquire non-hyperpolarized $^{19}F$ images for reference ($n_3 = 2000$);
e) acquire eight hyperpolarized FIDs ($n_4 = 8$);
f) acquire non-hyperpolarized $^{19}F$ FIDs for reference ($n_5 = 8000$).

The data presented in Fig. 5 of the manuscript were obtained after the experimental parameters for spectroscopy were changed to 16k bandwidth and 4k data points to increase the spectral resolution.

## 2. $^{19}F$ coil sensitivity

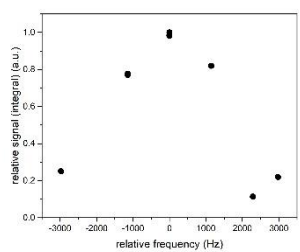

**SI Fig. 2**
Integrated values of the phase-corrected real part of the $^{19}F$ spectrum of pure trifluoroethanol (TFE) as a function of the difference between the transmit/receive frequency of the 90° pulse and the frequency of the maximum resonance of the TFE triplet (22,901,550 Hz). The integrals were scaled relative to the maximum value.

## 3. Simultaneous dual channel ¹⁹F/¹H spectroscopy

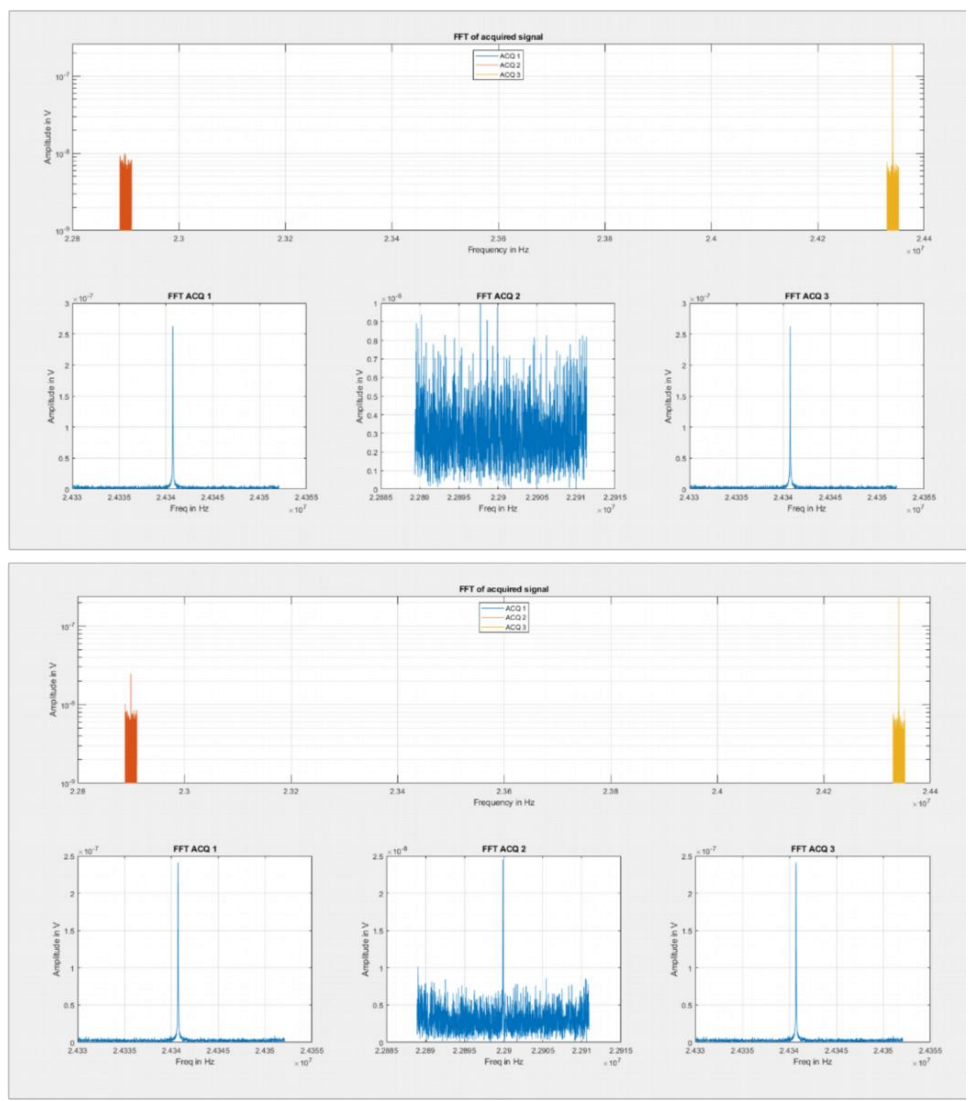

**SI Fig. 3:**
Graphical user interface (GUI) provided by the manufacturer for displaying of the acquired multichannel data.
Single scan spectra of a non-illuminated (upper panel) and illuminated (lower panel) solution of 1.943 mmol/L 3-fluoro-D/L-tyrosine/0.270 mmol/L riboflavin 5'-monophosphate in a $D_2O+H_2O_{dest.}$ solution (580 μL $D_2O$ + 20 μL $H_2O_{dest.}$). 2k data points were acquired with 22 kHz bandwidth. The upper part of each panel shows the simultaneously acquired $^{19}F$ (red) and $^1H$ (yellow) signals. The logarithmic scaling allows a quick estimation of the noise level and frequencies of each channel. The lower parts of each panel show three data acquisition channels (left: $^1H$; center: $^{19}F$; right: additional nucleus, here $^1H$). The GUI allows interactive use of *Matlab* graphical routines (zoom, pan, export etc.) to display information in more detail.
*Upper panel*: Without illumination, only the $^1H$ signal can be seen.
*Lower panel*: After 15 s of illumination, the $^{19}F$ signal is clearly visible. The SNR is still significantly lower than that of the $^1H$ data.

## 4. Decay of $^{19}$F and $^1$H FIDs

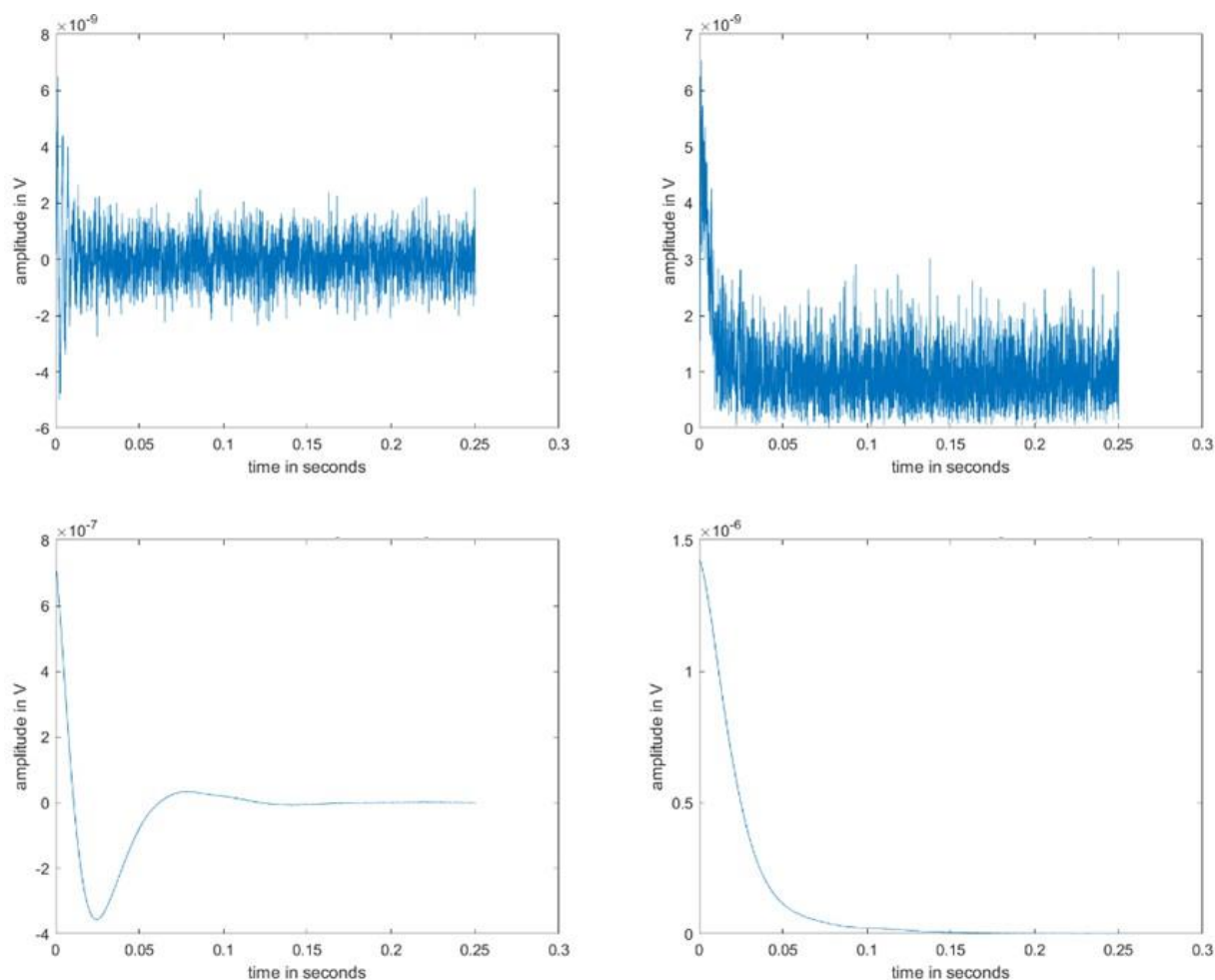

**SI Fig. 4**
Signal decay for $^{19}$F and $^1$H of 8000 averaged FIDs of a non-illuminated sample containing 1.943 mmol/L 3-fluoro-D/L-tyrosine and 0.270 mmol/L riboflavin 5'-monophosphate in $D_2O+H_2O_{dest.}$ solution (580 μL $D_2O$ + 20 μL $H_2O$).

*Upper left*: phase-corrected real part of 8000 averaged $^{19}$F FIDs.
*Upper right*: magnitude spectrum of 8000 averaged $^{19}$F FIDs.
*Lower left*: phase-corrected real part of 8000 averaged $^1$H FIDs.
*Lower right*: magnitude spectrum of 8000 averaged $^1$H FIDs.

The $^{19}$F signal decays to the baseline within about 30-50 ms, the $^1$H signal within about 150 ms.
The $^1$H signal originates from the protons of the $H_2O$ added to the solution to enable shimming on the $^1$H channel and $^1$H imaging. All data were acquired with 4k data points and 16 kHz bandwidth.